\documentclass[12pt]{JHEP3}
\pdfoutput=1

\usepackage{amssymb, amsmath, amsopn, amsthm}
\usepackage{epsfig}

\usepackage{graphicx}

 \def\makeatletter{\catcode`\@=11}
 \makeatletter
 \def\mathbox#1{\hbox{$\m@th#1$}}%
 \def\math@ccstyles#1#2#3#4#5#6#7{{\leavevmode
       \setbox0\mathbox{#6#7}%
       \setbox2\mathbox{#4#5}%
       \dimen@ #3%
       \baselineskip\z@\lineskiplimit#1\lineskip\z@
       \vbox{\ialign{##\crcr
              \hfil \kern #2\box2 \hfil\crcr
             \noalign{\kern\dimen@}%
              \hfil\box0\hfil\crcr}}}}
 \def\mathaccstyles{\math@ccstyles\maxdimen}
 \def\maththroughstyles{\math@ccstyles{-\maxdimen}}
 \def\unity%
    {\maththroughstyles{.45\ht0}\z@\displaystyle {\mathchar"006C}\displaystyle 1}

\title{ Interpolating from Bianchi Attractors to Lifshitz and  AdS Spacetimes } 
\author{Shamit Kachru$^1$, Nilay Kundu$^2$, Arpan Saha$^3$, Rickmoy Samanta$^2$ and Sandip~P.~Trivedi$^2$
~\\
$^1$SITP, Department of Physics and\\
Theory Group, SLAC \\
Stanford University\\
Stanford, CA 94305, USA \\
 ~\\
$^2$Tata Institute for Fundamental Research  \\
Mumbai 400005, India\\
~\\
$^3$Indian Institute of Technology -- Bombay \\
Powai, Mumbai, India \\

\vspace{0.05cm}

E-mail: \email{skachru@stanford.edu, nilay.tifr@gmail.com, arpansaha2007@gmail.com,  rickmoysamanta@gmail.com, trivedi.sp@gmail.com} \\
}

\abstract{ {\fontsize{0.36cm}{1em}\selectfont 
We construct classes of smooth  metrics which interpolate from Bianchi attractor
geometries  of Types II, III, VI and IX in the IR to Lifshitz or $AdS_2\times S^3$ geometries in the UV. While we do
 not obtain these metrics as solutions of Einstein gravity coupled to a simple matter field theory, we show that 
the matter sector stress-energy required to support these geometries (via the Einstein equations) does satisfy the weak, 
and therefore also the null, energy condition. Since  Lifshitz or $AdS_2\times S^3$ geometries can in turn be 
connected to $AdS_5$ spacetime,
our results show that there is no barrier, at least at the level of the energy conditions, for  solutions to arise connecting 
these Bianchi attractor geometries to $AdS_5$ spacetime. The asymptotic $AdS_5$ spacetime has no non-normalizable
 metric deformation turned on, which suggests that furthermore, the  Bianchi attractor geometries can be  the IR geometries 
dual to field theories living in flat space, with the breaking of  symmetries being either spontaneous or due to sources for 
other fields. Finally, we show that for a large class of flows which connect two Bianchi attractors, a C-function can be defined which is monotonically decreasing from the UV to the IR as long as the null energy condition is satisfied. However, except for special examples of Bianchi attractors (including  AdS space), this function does not attain a finite and non-vanishing constant value at the end points.  
}}

\preprint{ TIFR/TH/13-24}

\def\be{\begin{equation}}

\def\ee{\end{equation}}

\def\bea{\begin{eqnarray}}

\def\eea{\end{eqnarray}}

\begin{document}
 \setcounter{page}{2}

\section{Introduction}
Recent developments suggest that fascinating connections might exist between the study of gravity and 
 condensed matter physics, or more specifically the study of strongly coupled field theories at finite density.  See \cite{Hartnoll, Herzog:2009xv, McGreevy:2009xe, Sachdev:2011wg}
for reviews of this subject with additional references.

On the gravity side, motivated by the varied and beautiful phases found in nature, new brane solutions have been discovered. These branes have new kinds of hair, or have horizons with reduced symmetry. For example,  see \cite{Gubser:2008px, Hartnoll:2008vx, Hartnoll:2008kx, Horowitz:2010gk} for discussions on how black hole no-hair theorems can be violated in AdS space in the context of holographic superconductivity; \cite{lifsol1,lifsol2,lifsol3} for discussions of how emergent horizons with properties reflecting dynamical scaling in the dual field theory (``Lifshitz solutions") can arise; and \cite{hv1,hv2,hv31,hv3,hv32,hv4,hv5} for discussions of horizons exhibiting both dynamical scaling and
hyperscaling violation.\footnote{Embeddings of such solutions in string theory have also been discussed in many papers, such as \cite{hv4,hv5,BalNar,DG,Gregory,Narayan:2012hk,Singh:2012un,Dey:2012tg,Dey:2012rs}.}
The earliest work mostly focused on horizons with translational and rotational symmetry, but more recently examples of
black brane horizons dual to field theories with further reduced space-time symmetries have been  discussed in e.g. \cite{harvey,ooguri,gauntlett,IKKNST,IKKNSTW,Nori,Horowitz,Erdmenger,Rozali,DonosHartnoll,Cremonini,Vegh,Ning,Donos,
NingSarah,Sarah,Chesler}.

Extremal branes are particularly interesting, since they correspond to ground states of the dual 
field theory in the presence of a chemical potential or doping. Their near-horizon geometries  
 often exhibit a type of  attractor behavior, and as a result, are quite universal and independent of 
many details. There is a considerable body of work regarding the attractor mechanism, starting with the pioneering work in \cite{FKS}. A recent review with additional references is \cite{lectures}. For
studies of the attractor mechanism in the non-supersymmetric context, which is the situation of relevance here, see for instance \cite{Ga,Gibbons,Sen,attractors,Renata,KKS}.
 
Of particular interest for this paper are the brane solutions studied in \cite{IKKNST,IKKNSTW}, which correspond to phases of matter which are homogeneous but not isotropic.  
 It was shown that  in $4+1$ dimensions,  
 such brane solutions    can be classified using the Bianchi classification 
developed earlier for studying homogeneous cosmologies. 
In \cite{IKKNST,IKKNSTW}, it was found that for extremal black branes of this kind,  the  near-horizon geometry itself    often 
 arises as an  exact  solution   for a system consisting of Einstein gravity coupled to (simple) suitable matter in the presence of a negative cosmological constant. 
These near-horizon solutions were given the name ``Bianchi attractors''.

The attractor nature  mentioned above  makes the Bianchi attractor  geometries  
more universal, and  therefore in many ways more interesting, than the   
complete extremal black brane solutions 
from which they arise in the IR. However, some examples of  more complete solutions, 
 interpolating between asymptotically AdS space and Bianchi attractors of various Types, are well
 worth constructing  and could lead to   important insights. 

For example,  Bianchi attractors have a non-trivial geometry along the field theory directions.
It is therefore  worth asking whether these attractors  can arise in  situations where the dual field theory 
lives in flat space, as opposed to the more baroque possibility that the UV field theory itself must be  
 placed in a non-trivial geometry of the appropriate Bianchi Type. This question maps to constructing 
interpolating  extremal black brane solutions and asking whether  
 the non-normalizable deformations for the metric can be asymptotically turned off near the AdS boundary
which lies at the ultraviolet end.

 For one  case,  Bianchi Type VII, 
 an explicit interpolating solution of this type   was  indeed found in \cite{IKKNST}. More precisely, it was seen that, in the
 presence of  suitable matter, 
 a solution exists 
which interpolates between the Bianchi attractor region and $AdS_2 \times R^3$. The latter in turn is well 
known to arise as the near-horizon region of an extremal Reissner--Nordstrom black brane which is asymptotically  $AdS_5$.
In this way, it was shown that Bianchi Type VII can arise as the near-horizon limit of an asymptotically 
AdS brane. In this solution, no non-normalizable mode for the metric is  turned on near the $AdS_5$
 boundary, and therefore the field theory lives in flat $3+1$ dimensional spacetime. Sources are turned on for 
some of the field theory operators (but none dual to non-normalizable metric modes),
and these operators are responsible for the breaking of UV symmetries that leads to Bianchi Type VII.

For the other Bianchi classes, finding such interpolating extremal brane solutions has proved 
difficult so far. The main complication is a calculational one. It is easy to write down a continuous
and sufficiently  smooth metric  which interpolates between the near-horizon region 
and asymptotic AdS space,  with no non-normalizable metric deformations turned on, for any of the other Bianchi classes. But it is not easy to find such a metric as an explicit solution to the Einstein equations for 
gravity coupled to some simple matter field theory. 
The symmetries of Type VII are a subgroup 
of the three translations and the rotations in $R^3$;  this allows the equations for the full 
interpolating solution in the Type VII case  to be reduced to algebraic ones, and solved. On the other hand, the symmetries in the other Bianchi Types cannot be embedded in those of  $R^3$, and thus the equations cannot be reduced to merely algebraic ones.

Here,  we take a partial step towards finding such interpolating solutions for some of the other 
Bianchi classes.  We start with a  particular 
smoothly varying metric which interpolates between the near-horizon 
region and  Lifshitz spacetime. The metric is chosen so that the non-normalizable deformations of the metric near the Lifshitz
 boundary are turned off. While we do not obtain these metrics as solutions of Einstein gravity coupled to a specific simple matter field theory,
we demonstrate that were they to arise as  solutions, the required matter would satisfy the weak energy condition. 
In this way, we establish that there is no fundamental barrier, at least at the level of reasonable energy 
conditions, to having such an interpolating solution. 

In turn, it is well known that Lifshitz spacetimes, now thought of as the IR end,
 can  be connected to AdS space in the UV. Solutions of this type to  Einstein's equations coupled with  
reasonable matter satisfying the energy conditions   have been obtained, see, e.g., \cite{Braviner:2011kz}, \cite{Korovin:2013bua}, \cite{lifsol3}, \cite{Narayan:2012hk},
 \cite{Dey:2012tg}, \cite{Dey:2012rs}, \cite{Kumar:2012ui}, \cite{Singh:2013iba}, \cite{Faedo:2013aoa}.
 In these solutions often  
 no non-normalisable metric deformations are turned on in the AdS region, although a source for other operators can be 
  present. 
Taking these solutions  together with the interpolating metrics we study, one can then 
 conclude that interpolating geometries exist which 
connect some of the Bianchi classes to asymptotic AdS space. These interpolations do not 
 violate  the energy conditions, and they do not have any   non-normalisable deformations for the metric 
 turned on in the asymptotic AdS region. 
This establishes one of the main results of our paper. 


Hopefully, our result will provide motivation for finding  solutions of Einstein's equations sourced by 
suitable specific matter field theories, which interpolate between the Bianchi classes and Lifshitz or  AdS spaces, in the near future. 
 The  weak energy condition implies the null energy condition. Thus, our results also imply that no violations of the null energy condition are necessary for the required interpolations. 
While violations of the null-energy condition are known 
to be possible, they usually 
require either quantum effects or exotic objects like orientifold planes in string theory.  
Our result suggests that these are not required, and that standard matter fields   should suffice as sources in constructing these interpolating solutions. 
Once constructed,   these solutions  will allow us to ask whether, from the field theory perspective, 
 the  symmetries of various Bianchi classes can emerge in the IR, either spontaneously or in response to 
some suitable source not including the metric. 

Near the end of the paper, in \S6, we also explore the existence of C-functions in flows between Bianchi attractors. This section can be read independently of the earlier parts of the paper.
We find that if the matter  sourcing the geometry satisfies the null energy condition,
a function does exist, for a large class of flows, which is monotonically decreasing from the UV to the IR. But unless the attractors
meet a special condition, this function does not attain a finite, non-vanishing constant value at 
the end points. We also show that the area element of the three-dimensional submanifold generated by the 
Bianchi isometries in the attractor spacetimes monotonically decreases from the UV to the IR.

The plan of the paper is as follows. In \S2, we discuss the weak and null energy conditions. \S3 outlines the general procedure we follow in constructing the interpolating metrics and illustrates this for
 the particular case of Bianchi Type II. Bianchi Type VI and the closely related classes of Type III and V are discussed in \S4, and Type IX, for which the interpolation is to $AdS_2\times S^3$, is discussed in \S5. In \S6, we explore the existence of a C-function.  We end with some conclusions in \S7. The appendix contains a more complete discussion of the energy conditions.

\section{Energy Conditions }
We will work in $4+1$ dimensional spacetime and adopt the mostly positive convention for the 
metric, so that the flat metric is $\eta_{\mu\nu}=\mathrm{diag}(-1,1,1,1,1)$.   

Consider a coordinate system $x^\mu,\, \mu=0, 1, \ldots, d$, with the metric being $g_{\mu\nu}$. 
We denote the stress energy tensor, as  in the standard notation, by  $T_{\mu\nu}$, and let $n_\mu$ 
be a null vector, with $n_\mu n_\nu g^{\mu\nu}=0$.
Then the  null energy condition (NEC)  is satisfied iff
\be
\label{nullene1}
T_{\mu\nu} n^\mu n^\nu \ge 0
\ee
for any future directed null vector, see \cite{Poisson}, \cite{Hawking-Ellis}.
Here we will only consider spacetimes which are time reversal invariant, i.e., with a $t\rightarrow -t$ symmetry. For such spacetimes the requirement of $n^\mu$ being future directed can be dropped. 

For the purposes of our analysis it is convenient to state  this condition as follows. 
$T^\mu_\nu$ can be regarded as a linear operator acting on tangent vectors. Let the orthonormal eigenvectors
of this operator be denoted by $\{u_0, u_1,u_2,u_3,u_4\}$, with eigenvalues, $\{\lambda_0, \lambda_1,\lambda_2,\lambda_3,\lambda_4\}$ respectively. Note that orthonormality implies $\langle u_a,u_b\rangle \equiv u_{a\mu} u_{b\nu} g^{\mu\nu}=\eta_{ab}$, so that $u_0$ is time-like and the other eigenvectors, $u_c,\, c=1,\ldots, 4$, are space-like. 

Then, as discussed in Appendix A, the NEC requires that 
\be
\label{nec2}
-\lambda_0+\lambda_c \ge 0
\ee
for $c=1,2,3,4$. 

In contrast, the weak energy condition (WEC) requires that  
\be
\label{wec1}
T_{\mu\nu} u^\mu u^\nu \ge 0,
\ee
for any future directed time-like vector $u^\mu$ \cite{Poisson}, \cite{Hawking-Ellis}. 
As in the discussion of the NEC above, for the time reversal invariant backgrounds we consider here, 
the requirement that $u^\mu$ is future directed need not be imposed. 
In terms of the eigenvalues $\{\lambda_0, \lambda_c\}$ of $T^\mu_\nu$, 
this leads to two conditions: 
\begin{align}
\lambda_0  & \le   0  \label{wec2a} \\
\lambda _c -\lambda_0 &\ge 0,\quad \mbox{for } c = 1,2,3,4 \label{wec2b} .
\end{align}
 
From eq.(\ref{wec2b}) and eq.(\ref{nec2}) we see that  
 the weak energy condition implies the null energy condition.  Thus, the weak energy condition is stronger. 

We make two  final comments before we end  this section. 
In this paper, we will follow the conventions of \cite{IKKNST}, where the action takes the form (see equation (3.4) of \cite{IKKNST})
\be
\label{act1}
S=\int d^5x\sqrt{-g}\,\{R + \Lambda + \cdots\}.
\ee
The ellipsis on the RHS denotes the contribution to the action from matter fields. 
In these conventions, $AdS_5$  spacetime is a solution to the Einstein equations,
 in the absence of matter, for $\Lambda>0$. It follows then that the cosmological constant required 
for AdS space violates eq.(\ref{wec2a}) and thus the weak energy condition, but it satisfies 
eq.(\ref{nec2}) as an   equality, thereby meeting  the null energy condition. 

 Secondly, we have assumed above that the linear operator $T^\mu_\nu$ 
is diagonalizable and that its eigenvalues are real. These properties do not have to be true, 
since $T^\mu_\nu$, unlike, $T_{\mu\nu}$, need not be symmetric, and moreover since the inner product 
is Lorentzian (see \cite{Stephani}).
However, for the interpolations we consider, it will turn out  that $T^\mu_\nu$ is indeed diagonalizable
 with real eigenvalues and therefore we will not have to consider this more general possibility. 

\section{Outline Of Procedure}
In this section, we will outline the basic ideas that we  follow to find   metrics with the required 
properties that interpolate between the near-horizon attractor region and an 
 asymptotic Lifshitz spacetime. We will illustrate this  procedure in the context of one concrete example, which we will  take to be 
Bianchi Type II.  Holography in this particular Bianchi attractor was recently studied in depth in \cite{Sarah}.

The metrics we consider in general have the form
\be
\label{metgena}
ds^2=-g_{tt}(r)dt^2+g_{rr}(r) dr^2 + \sum_{i,j=1,2,3} g_{ij}(r,x^i) dx^idx^j.
\ee
In the Bianchi attractor region which occurs in the deep  IR,  for $r\rightarrow -\infty$, the metric takes the form,
\be
\label{metatt}
ds^2_{B}=-e^{2\beta_tr} dt^2 + dr^2 + \sum_{i=1,2,3} e^{2\beta_i r} (\omega^i)^2,
\ee
where $\omega^i$ are one-forms invariant under the Bianchi symmetries generated by the Killing fields $\xi_i$, $i=1,2,3$
(More generally, off-diagonal terms are also allowed in eq.(\ref{metatt}) but we will not consider this possibility here.)
The commutation relations of the Killing vectors
\footnote{The Bianchi classification is described in \cite{LL}, \cite{SR}, including the symmetry generators and invariant one-forms;
also see Section 2 and Appendix A of \cite{IKKNST} for a general discussion of this classification more akin to our purpose.}
\be
\label{balga}
[\xi_i,\xi_j]=C^k_{ij}\xi_k
\ee
 give rise to the corresponding Bianchi algebra.

In the  far UV on the other hand, which occurs for $r \rightarrow \infty$, the metric becomes of Lifshitz form,
\be
\label{metlif}
ds_L^2= -e^{2{\tilde \beta}_t r} dt^2 +  dr^2 + e^{2{\tilde \beta} r} \sum_{i=1,2,3} dx_i^2.
\ee
Here for simplicity, we only consider the case where all the spatial directions have the same scaling exponent,
 ${\tilde \beta}$, more generally this exponent can be different for the different spatial directions. 
Also, to avoid unnecessary complications  we take the exponent in the time direction ${\tilde \beta}_t$ in the Lifshitz region to satisfy the condition 
\be
\label{condouta}
{\tilde \beta}_t =\beta_t,
\ee
where $\beta_t$ is the value for the exponent in the Bianchi attractor region, eq.(\ref{metatt}). The metric eq.(\ref{metlif}) then becomes
\be
\label{metbou}
ds^2_{L}=-e^{2\beta_tr} dt^2 + dr^2 + e^{2\tilde\beta r}\sum_{i=1,2,3}  (dx^i)^2.
\ee

The metric which interpolates between these two regions is taken to have the form
\be
\label{metgen}
ds^2=\left(\frac{1-\tanh \sigma r}{2}\right)ds_{B}^2+\left(\frac{1+\tanh\sigma r}{2}\right)ds_L^2,
\ee
where $ds_{B}^2$ and $ds_L^2$ are defined in eq.(\ref{metatt}) and eq.(\ref{metlif}) respectively. 
$\sigma$ is a positive constant which characterizes how rapid or gradual the interpolation is.  
One can show, and this will become clearer in the specific examples we consider below,   
  that as long as $\sigma$ is sufficiently big 
the metric becomes of the Bianchi attractor form as $r\rightarrow -\infty$. 
Also, for sufficiently large $\sigma$ the metric becomes of Lifshitz type as $r\rightarrow \infty$.
More correctly, for this latter statement to be true  the limit $r\rightarrow \infty$ must be taken keeping 
 the spatial coordinates $x^i, i=1,2,3$  fixed. We will also comment on this order of limits in
 more detail below.

We should emphasize that  we do not obtain  the interpolating metric in eq.(\ref{metgen}) 
as a
 solution to Einstein's equations coupled to suitable matter. Instead, what we will do is to construct from the metric,
via the Einstein equations, a stress energy tensor for matter and then examine whether this stress energy 
 satisfies the energy conditions. 

Below, we will analyze  cases where the interpolation is from attractor geometries of Bianchi
Type  II, III, V, or VI to 
Lifshitz geometry. In addition, using a different strategy,  we will also consider the interpolation from Type IX to $AdS_2\times S^3$.
 
 \subsection{More Details for the Type II Case}
 Let us now give more details for how the analysis proceeds in the Type II case.
 
It will be convenient in the analysis to take the  Bianchi attractor geometry 
 and the Lifshitz geometry  which arise in the IR and UV ends of the interpolation  as solutions of Einstein's equations coupled to reasonable matter. This ensures that the energy conditions will be satisfied at least asymptotically. 
In fact the Bianchi attractor geometry and the Lifshitz geometry can both arise as solutions in a system of gravity coupled to a massive Abelian 
gauge field in the presence of a cosmological constant, with an action of the 
 form,
\be
\label{actionII}
S= \int d^{5}x \sqrt{-g} \left(R +\Lambda  -\frac{1}{4} F^{2}-\frac{1}{4}m^{2}A^{2}\right).
\ee

The Type II solutions which arise from this action were discussed in \cite{IKKNST} and we will mostly follow the same 
 conventions here. 
 The invariant one-forms for Type II are given by
 \begin{equation}
 \label{invfII}
 \omega^1 =  dy-x\,dz,\quad
 \omega^2 =  dz,\quad
 \omega^3 = dx.
 \end{equation}
 The   solutions of Type II obtained from eq.(\ref{actionII})  were described in eq.(4.2), (4.3)  and (4.10), (4.11) 
 in \cite{IKKNST}. 
 The metric and gauge field in these solutions  take the form 
 \be
 \label{intmet}
 ds_{B}^2 = R^2[dr^2 - e^{2\beta_t r}dt^2 +e^{2(\beta_2 +\beta_3)r}(\omega^1)^2+e^{2\beta_2r}(\omega^2)^2+
e^{2\beta_3r}(\omega^3)^2]
 \ee
  and \be
  \label{gfII}
 A = \sqrt{A_t}\,e^{\beta_t r}dt.
 \ee
 These   solutions are  characterized by five parameters,  $ R, \beta_t,\,\beta_2,\,\beta_3,\, A_t$.
The equations of motion give rise to five independent equations which determine these parameters in terms 
of $m^2, \Lambda$.  
 For our purposes it will be convenient to work in units where $R=1$ and to use the equations of motion to express $\beta_t,\,\beta_1,\,\beta_2,\, A_t$ and $m^2$ in terms of $\Lambda$.
 The resulting relations are,
 \begin{align}
 \label{btII}
 \beta_t &= v,\\
 \label{b2II}
 \beta_2=\beta_3 &= -\frac{(3  - \Lambda + 
       u)v}{36 - 8 \Lambda},\\
 \label{mII}
       m^{2}&=\frac{8}{11}(6  -\Lambda + u),\\
 \label{atII}
       A_t&=\frac{-11v^2+3u}{18-4 \Lambda },
 \end{align}
 where
 \begin{equation*}
 \begin{split}
 u&=\sqrt{-63  + 
    10  \Lambda + \Lambda^2},\\
 v&=\left[\frac{-81  + 19 \Lambda + 
  3 u}{22}\right]^{\frac{1}{2}}.
  \end{split}
 \end{equation*} 
Demanding that $ A_{t},m^{2},\beta_{t}, \beta_2,\beta_3 $ be positive and $u$ be real, we get $\Lambda > \frac{9}{2}$.
The Lifshitz metric which we are interested in  near the boundary also arises  as a solution from the 
 action in eq.\eqref{actionII}.   The metric and gauge field in this solution  take the form 
 \be
 \label{Lifsection2}
 ds_L^2 = dr^2 - e^{2\beta_t r}dt^2 +e^{2\tilde \beta r}dx^2+e^{2\tilde \beta r}dy^2+e^{2\tilde \beta r}dz^2
 \ee
  and \be
  \label{gfLif}
 A = \sqrt{A_t}\,e^{\beta_t r}dt.
 \ee
 The   solution is characterized by three parameters, $\beta_t,\,\tilde\beta, A_t$
 which are determined in terms of $m^2$ and $\Lambda$. 
 For our purposes it is more convenient to express $\tilde\beta, A_t$ and $m^2$ in terms of $\beta_t$ and $\Lambda$. 
 These relations, which arise due to the equations of motion, are
 \begin{align}
 \label{bLif}
 \tilde\beta &= \frac{1}{9}\left(-\beta_t+ \sqrt{-8\beta_t^2+9\Lambda}\right),\\
 \label{mLif}
       m^{2}&=\frac{2}{3}\beta_t\left(-\beta_t+ \sqrt{-8\beta_t^2+9\Lambda}\right),\\
 \label{atLif}
       A_t&=\frac{2}{9}\left(10 - \frac{1}{\beta_t}\sqrt{-8\beta_t^2+9\Lambda}\right).
 \end{align}
In order to ensure that $\tilde\beta,A_t,m^2$ are all nonnegative, we must have 
$\beta_t>0$, $\beta^2_t \le \Lambda \le 12\beta_t^2$. We will consider Lifshitz metric where these conditions hold. 

The Type II and Lifshitz solutions we consider correspond to the same value of the cosmological constant.
It will also be convenient to take the exponent $\beta_t$ along the time direction in the Type II and 
Lifshitz cases to be the same as discussed in eq.(\ref{condouta}). This will mean that the mass parameter $m^2$
 for the Type II and Lifshitz cases will be different in general.
 
A negative cosmological constant (in our conventions $\Lambda>0 $ )  violates the weak energy condition, thus in studying the 
 violations of this condition it is useful to separate the contributions of the cosmological constant from the matter in the stress energy. Since the two asymptotic geometries we consider arise as 
solutions with the same value of the cosmological constant we can consistently take the 
cosmological constant to have  this same  value   throughout  the interpolation. Using the Einstein equations we 
 can then define 
  a matter stress tensor, minus the cosmological constant, and then study its behavior  with respect to the 
 weak energy condition.   The null energy condition, in contrast to the weak energy condition, does not receive contributions 
from the  cosmological constant, and so for studying its possible violations such a separation between matter and the cosmological constant components  is not necessary.

We now turn to the full interpolating metric. 
As discussed in the previous subsection this takes  the form
\begin{equation}
\label{intmetII}
\begin{split}
ds^{2}&=dr^{2}- e^{2\beta_{t} r }dt^{2}\\
&\quad+\left[\left(\frac{1-\tanh\sigma  r }{2}\right)e^{2\beta_{3} r }+\left(\frac{1+\tanh \sigma r }{2}\right)e^{2\tilde\beta r }\right] dx^{2}\\
&\quad+\left[\left(\frac{1-\tanh\sigma  r }{2}\right)e^{2(\beta_{2}+\beta_{3}) r }+\left(\frac{1+\tanh\sigma  r }{2}\right)e^{2 \tilde\beta r }\right] dy^{2}\\
&\quad+ \left[\left(\frac{1-\tanh\sigma  r }{2}\right)(x^{2}e^{2(\beta_{2}+\beta_{3}) r } +e^{2\beta_{2} r })+\left(\frac{1+\tanh \sigma r }{2}\right)e^{2\tilde{\beta} r }\right]dz^{2}\\
&\quad-x\left(\frac{1-\tanh \sigma r }{2}\right) e^{2(\beta_{2}+\beta_{3}) r }(dy\otimes dz +dz\otimes dy).
\end{split}
\end{equation}
We note that in the limit of $r$ becoming very large, the above may be approximated by 
 \begin{equation}
\label{intmetIIapprox}
\begin{split}
ds^{2}&=dr^{2}- e^{2\beta_{t} r }dt^{2}+\left[e^{2(\beta_{3}-\sigma) r }+e^{2\tilde\beta r }\right] dx^{2}\\
&\quad+\left[e^{2(\beta_{2}+\beta_{3}-\sigma) r }+e^{2 \tilde\beta r }\right] dy^{2}\\
&\quad+\left[x^{2}e^{2(\beta_{2}+\beta_{3}-\sigma) r } +e^{2(\beta_{2}-\sigma) r }+e^{2\tilde{\beta} r }\right]dz^{2}\\
&\quad-x e^{2(\beta_{2}+\beta_{3}-\sigma) r }(dy\otimes dz +dz\otimes dy).
\end{split}
\end{equation}
To ensure that this metric approaches the  Lifshitz geometry as $r\rightarrow \infty$,
  with exponentially small corrections,  the terms arising from the Lifshitz metric, eq(\ref{Lifsection2}),
 must dominate in every component of the metric. 
It is easy to see that this condition is met when  
\be
\label{condass2}
\sigma > \beta_2 + \beta_3.
\ee
Similarly, one finds that the conditions requiring the metric to become of the Bianchi II type, eq.(\ref{intmet}), in the IR are also met when $\sigma$ satisfies the condition in eq.(\ref{condass2}). 

Actually, the $r\rightarrow +\infty$ limit is a bit subtle. As one can see from
 the coefficient of the $dz^2$ and the $(dy\otimes dz +dz\otimes dy)$ terms   in 
eq.(\ref{intmetIIapprox}),
eq.(\ref{condass2}) ensures that the metric becomes of Lifshitz type when $r\rightarrow \infty$, 
as long as   $x$ is constant,  or at least for $|x|$ growing sufficiently slowly in this limit.
This is in fact the limit we will consider in our discussion. 

Taking the limit in this way is  well motivated physically. It is quite reasonable to place the dual field theory 
whose properties we are interested in studying in a box of finite size. 
In fact this is  always the case in any experimental situation. In such a finite  box the   
range of the spatial coordinates is finite ensuring that the   $r\rightarrow \infty$ limit
 is of the required type.  As long as the box is sufficiently big,
compared to the other scales, e.g. the temperature, the properties of the system, e.g. its thermodynamics, 
do not depend in a sensitive way on the size of the box. 

While the requirement for getting the correct asymptotic behavior imposes a lower bound on $\sigma$, eq.(\ref{condass2}),  
meeting  the energy conditions give rise  to an upper bound on 
$\sigma$, as we will see below. It will turn out that there is a finite 
 region for the allowed values of $\sigma$ between  these two bounds, for the Type II case, and by choosing $\sigma$ to lie in
 this region an acceptable interpolation meeting the energy conditions can be obtained.

%

 \subsubsection{Energy Conditions for the  Type II Interpolation}
 With the interpolating metric in hand, we can now test the various energy conditions. 
 We do so numerically. 

  From the metric, eq.(\ref{intmetII}),
  we define  a  stress tensor,  assuming that the Einstein equations are valid. This gives
 \be
 \label{defstress}
 T_{\mu\nu}\equiv   R_{\mu\nu}-{1\over 2} g_{\mu\nu} R.
 \ee
 (We set $\kappa=8 \pi G_N=1$.)
 This stress energy tensor in turn is  separated  into a matter and a cosmological constant 
 contribution. With our conventions, eq.(\ref{act1}), we get  
 \be
 \label{sepstress}
 T_{\mu\nu}={\Lambda \over 2}  g_{\mu\nu} + T^{(\mathrm{matter})}_{\mu\nu}.
 \ee
 Combining these two equations gives 
 \be
 \label{fmatterst}
 T_{\mu\nu}^{(\mathrm{matter})}= R_{\mu\nu}-{1\over 2} g_{\mu\nu} R -{\Lambda\over 2} g_{\mu\nu}.
 \ee
 To analyze whether the  energy conditions are valid, we first note that 
 owing to the form we have chosen for the interpolating metric, eq.(\ref{intmetII}),  
$T^{(\mathrm{matter}) \mu}_\nu$ is block diagonal. Therefore, its eigenvalues  
take the simple form
\begin{align}
\lambda_0 &= -{\Lambda\over 2}+ T_{t}^{ t}\label{eig0},\\
\lambda_1 &= {1\over 2}\left[-\Lambda+T_{r}^{ r}+T_{x}^{x}+\left[\left(T_{r}^{ r}-T_{x}^{ x}\right)^2+4T_{r}^{ x}T_{x}^{ r}\right]^{\frac{1}{2}}\right]\label{eig1},\\
\lambda_2 &= {1\over 2}\left[-\Lambda+T_{r}^{ r}+T_{x}^{ x}-\left[\left(T_{r}^{ r}-T_{x}^{ x}\right)^2+4T_{r}^{ x}T_{x}^{ r}\right]^{\frac{1}{2}}\right]\label{eig2},\\
\lambda_3 &= {1\over 2}\left[-\Lambda+T_{y}^{ y}+T_{z}^{ z}+\left[\left(T_{y}^{ y}-T_{z}^{ z}\right)^2+4T_{y}^{z}T_{z}^{ y}\right]^{\frac{1}{2}}\right]\label{eig3},\\
\lambda_4 &= {1\over 2}\left[-\Lambda+T_{y}^{y}+T_{z}^{ z}-\left[\left(T_{y}^{ y}-T_{z}^{ z}\right)^2+4T_{y}^{ z}T_{z}^{y}\right]^{\frac{1}{2}}\right]\label{eig4}.
\end{align}
Since we obviously have $\lambda_1\ge\lambda_2$ and $\lambda_3\ge\lambda_4$, the criteria  discussed in \S2 above reduces to just checking whether the following conditions hold: 
\be
\label{condass3}
\lambda_0\le0, \lambda_2-\lambda_0\ge 0,\lambda_4-\lambda_0\ge 0.
\ee
 
%


 For the numerics, we set  
 \be
 \label{vallam}
 \Lambda = 12.
 \ee
(In $R=1$ units). 

From eq.(\ref{b2II}) we can now determine $\beta_2, \beta_3$ and thus the lower bound on $\sigma$, eq.(\ref{condouta}), which turns
out to be $\sigma_{\mathrm{lower}}=0.5065$. 
As we increase $\sigma$ we  find in the numerical analysis 
that violations of the null energy condition start setting in around $\sigma= 1.05026$. The weak energy condition is not violated before this. Thus, there 
is a finite interval $0.5065<\sigma<1.05$,  within which both the correct asymptotic behavior for the metric  is obtained 
and the  null and weak energy conditions  are  met. 

To illustrate this, we consider the case where $\sigma=1$ in more detail. 
It turns out that $\lambda_2<\lambda_4$, where the eigenvalues are defined in eq.(\ref{eig0}),
eq.(\ref{eig1}),  eq.(\ref{eig2}), eq.(\ref{eig3}), eq.(\ref{eig4}). 

\begin{figure}
 \begin{center}
 \includegraphics[width=1\textwidth]{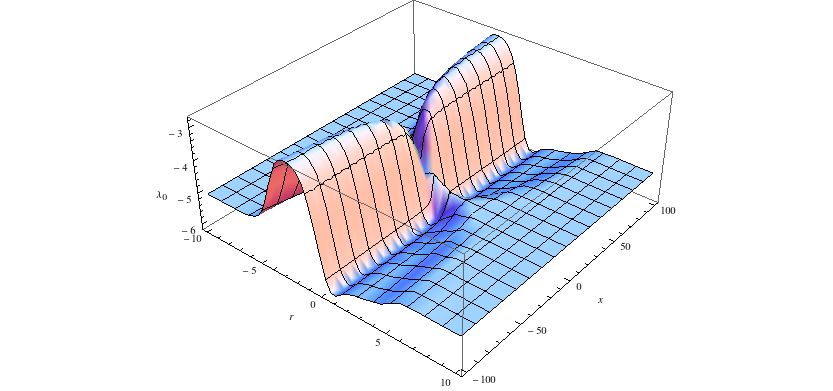}
 \caption{Type II 3D plot of  $\lambda_0$ (time-like eigenvalue) versus $r$ and $x$    for $\sigma=1,~\Lambda=12$.}\label{2weak}
 \end{center}
 \end{figure}

\begin{figure}
 \begin{center}
 \includegraphics[width=1\textwidth]{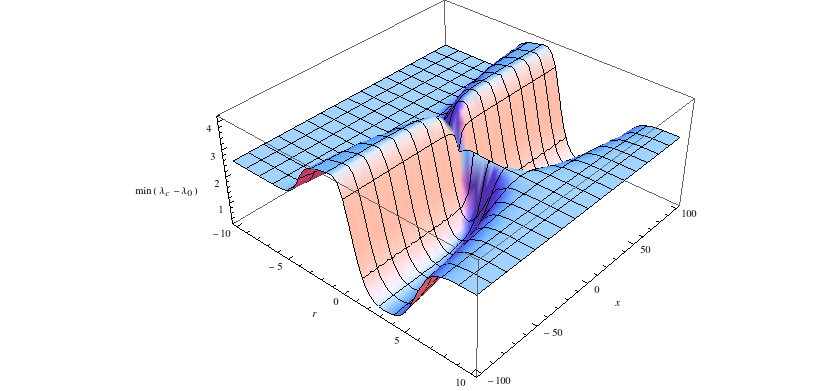}
 \caption{Type II 3D plot of   $\min(\lambda_{c}-\lambda_0)$ versus $r$ and $x$    for $\sigma=1,~\Lambda=12$.}\label{2null}
 \end{center}
 \end{figure}

\begin{figure}
 \begin{center}
 \includegraphics[width=0.7\textwidth]{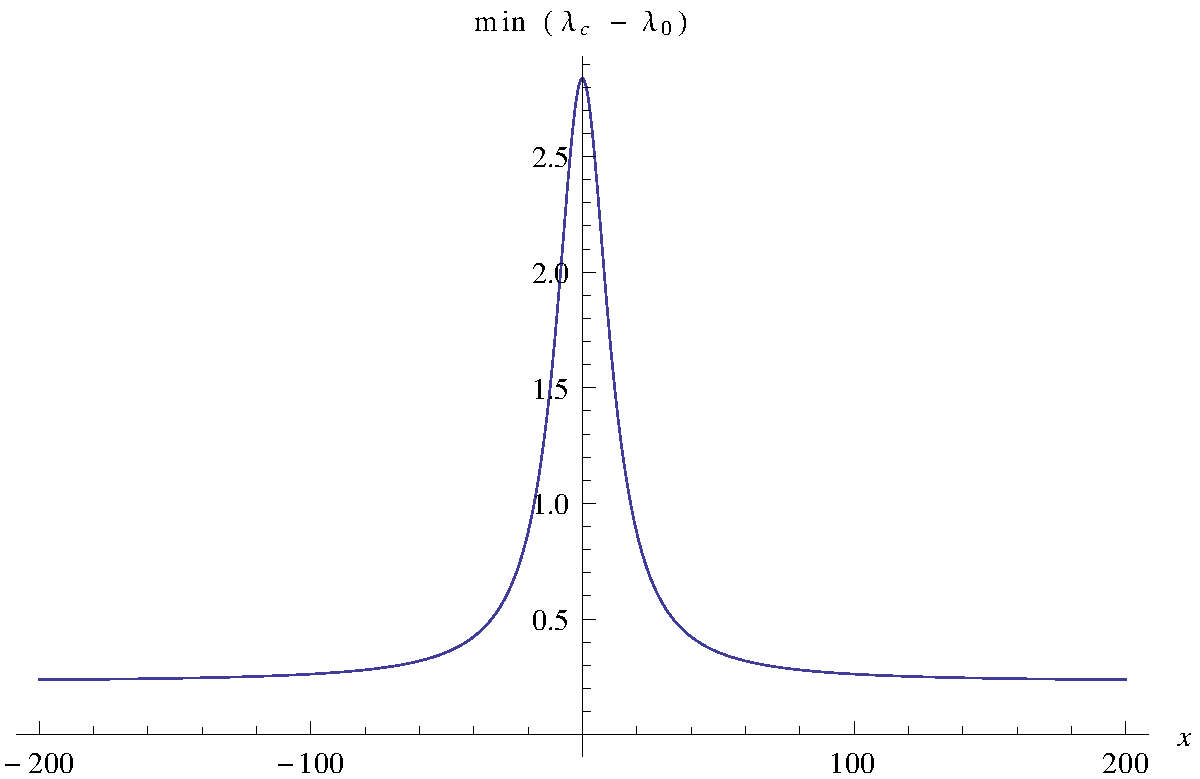}
 \caption{Type II plot of  $\min(\lambda_{c}-\lambda_0)$ versus $x$ at $r=3$ for   $\sigma=1,~\Lambda=12$.}\label{IIr3xprofile}
 \end{center}
 \end{figure}

\begin{figure}
 \begin{center}
 \includegraphics[width=0.7\textwidth]{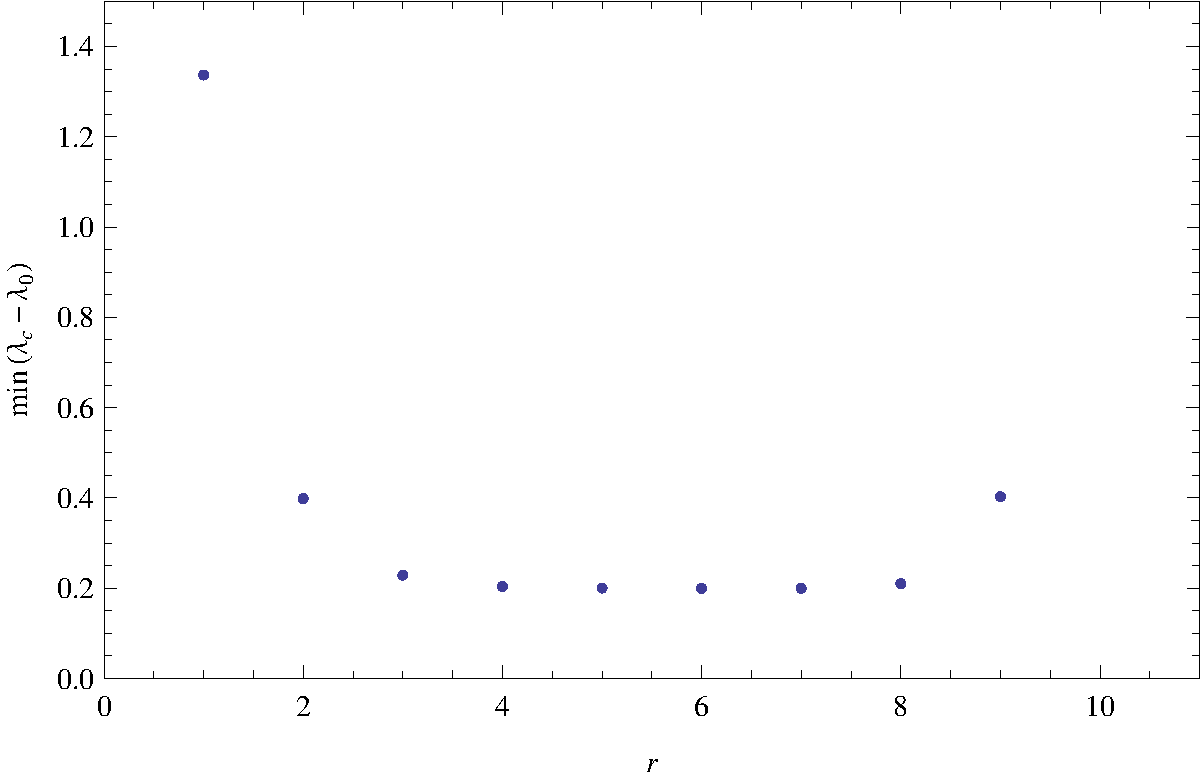}
 \caption{Type II list plot of  $\min(\lambda_{c}-\lambda_0)$ versus $r$ as $x\rightarrow\pm\infty$ for    for $\sigma=1,~\Lambda=12$.}\label{IIxinf}
 \end{center}
 \end{figure}

The plots of $\lambda_0$ and $\min(\lambda_c-\lambda_0)=\lambda_2-\lambda_0$, 
 are given in fig. \ref{2weak}, \ref{2null}. 
From fig. \ref{2weak} we see that $\lambda_0$ is always negative. 
In  fig. \ref{2null} we see that $\min (\lambda_c-\lambda_0) >0$ but there is a region around 
$r\sim 3$ where it  becomes very small. 
We have investigated this region further in much more detail numerically  and find that even after 
going out to arbitrarily large values of $x$,  $\min(\lambda_c-\lambda_0)$ continues to be positive in the
 worrisome range $2<r<8$. For a fixed value of $r$, in this range, 
 as we go out to larger $x$ the value of $\min(\lambda_c-\lambda_0)$ decreases reaching a minimum value for 
$|x| \rightarrow \infty$. For example, the resulting plot for $r=3$, as a function of $x$, 
is given in fig. \ref{IIr3xprofile} where we see that the minimum value obtained for $\min(\lambda_c-\lambda_0) $ is
 positive. For other values of $r$ in this range a qualitatively similar plot is obtained as $x$ is varied
with the minimum value of $\min(\lambda_c-\lambda_0)$ again being positive. 
As an additional check,  we have analytically computed the value of $\min(\lambda_c-\lambda_0)$ in the 
limit where  $|x|\rightarrow\infty$. In the worrisome region $2<r<8$ we find that this value is positive. 
We show this in fig. \ref{IIxinf}  where the limiting value of $\min(\lambda_c-\lambda_0)$, as $|x|\rightarrow \infty$,
 is plotted as a function of $r$.  We see that as $r$ increases, this limiting value  at 
first decreases,  reaching
 a minimum at around $r = 5$,  and then increases again. The minimum value is clearly positive showing that the null energy condition is indeed met everywhere in the interpolating metric.

  
Let us end this section with one comment. 
Because of the upper bound on $\sigma$, which arises in order to meet the energy conditions, the metric cannot approach
that of   Lifshitz space arbitrarily rapidly. 
The reader might worry that  the values of $\sigma$ allowed by this bound are too small  to be physically acceptable. 
To explain this, consider as an example the more familiar case of asymptotically $AdS_5$ spacetime.
  Since a domain wall in $AdS_5$
ought to carry positive energy density and pressure,
one might expect that the rate at which the metric of such a solution
  approaches  $AdS_5$  is  governed by   the normalizable metric deformations of $AdS_5$,
 and should not be slower.
 A similar type of argument should  also apply to Lifshitz spaces.
However, this expectation need not be valid if other fields are also  turned on, since
these fields can   source the metric, and this can   lead
 to a  fall-off slower than   that  expected from the normalizable mode of the metric itself.

\section{Types VI, V and  III }
We now turn to constructing metrics which interpolate from Bianchi Types VI,  III and V
to Lifshitz. Since our discussion will closely parallel that for Type II above, we will 
skip some details.  We will find that an analysis along the lines above will successfully lead to a class
of interpolating metrics for Type VI 
and Type III, meeting the weak and null energy conditions. However, for reasons which will become clearer below,
we do not succeed in finding such an interpolating metric for  Type V.

The algebra for a general Type VI spacetime is characterized by one parameter `$h$'.
 Killing vectors and invariant one-forms for Type VI are given in Appendix A of \cite{IKKNST} (see also \cite{SR}),
\be
\label{kvVI}
\xi_1=\partial_y,\quad \xi_2=\partial_z,\quad \xi_3=\partial_x+y\partial_y+h z \partial_z
\ee
and 
\be
\label{invfVI}
\omega^1=e^{-x}dy,\quad \omega^2=e^{-hx} dz,\quad \omega^3=dx~.
\ee
These depend on the parameter $h$. 

The Type V algebra is a special case of Type VI, and is obtained by setting $h=1$. The Killing vectors and invariant one-forms can then be obtained from eq.(\ref{kvVI}) and eq.(\ref{invfVI}) by setting $h=1$. Similarly the Type III algebra is also a special case obtained by setting $h=0$, 
with  the Killing vectors and one-forms  given by setting $h=0$ in the equations above. 

To keep the discussion simple, we restrict ourselves to only considering
  the case $h=-1$ for Type VI, besides also considering  the Type V and 
Type III cases.  

The invariant one-forms  for Type VI with $h=-1$ are 
\be
\label{infc}
\omega^1=e^{-x} dy,\quad \omega^2=e^{x} dz,\quad \omega^3=dx.
\ee

Bianchi Type VI attractor solutions, for the case $h=-1$, were obtained in Section 4.2 of \cite{IKKNST}
for a system of gravity coupled with a massive gauge field, with an action eq.(\ref{actionII}).
The solution  
  has a metric, 
\be
 \label{metVIa}
 ds_{B}^2 = R^2[dr^2 - e^{2\beta_t r}dt^2 +e^{2\beta_1 r}(\omega^1)^2+e^{2\beta_2r}(\omega^2)^2+
e^{2\beta_3r}(\omega^3)^2]
 \ee
 and a gauge field, eq.(\ref{gfII}), 
 with  the invariant one-forms  being given in eq.(\ref{infc}). As in the discussion for Type II we will work in $R=1$ units below.  
The exponents $\beta_t,\,\beta_1,\,\beta_2,\beta_3$ in the solution are then   given in terms of  $\Lambda$ by 
\begin{align}
\beta_t &=  v \label{btVI},\\
\beta_1 &=\beta_2 =  \frac{(-4+\Lambda-u)v}{24 - 4\Lambda} \label{b1VI} ,\\
\beta_3 & = 0,
\end{align}
while  the mass  and $A_t$ are
\begin{align}
m^{2}& =  \frac{2}{3}(8  -\Lambda + u) \label{mTypeVI},\\
A_{t} & =  \frac{-3v^2 + u }{6  - \Lambda }, \label{atTYpeVI}
\end{align}
where 
\begin{align}
u&=  \sqrt{-80  + 
 8  \Lambda + \Lambda^2}\label{defuVI}, \\
v&=\left[\frac{-28  + 5 \Lambda + 
  u}{6}\right]^{\frac{1}{2}} \label{defvVI}.
\end{align}
Demanding that $ A_{t},m^{2},\beta_{t}, \beta_1,\beta_2 $ be positive and $u$ be real, we get $\Lambda > 6$.
The Lifshitz spacetime in the UV is also obtained as a solution of the same system, eq.(\ref{actionII}).
The metric is given by eq.(\ref{Lifsection2}) and the gauge field by eq.(\ref{gfLif}). The exponent $\beta_t,{\tilde \beta}$ and $A_t$ are given in eq.(\ref {bLif}), (\ref{mLif}) and (\ref{atLif}) in terms of $m^2,\Lambda$. 
We will take the value of $\Lambda$ to be the same in the IR Type VI and the UV Lifshitz theories. 
For simplicity we will also take condition eq.(\ref{condouta}) to hold so that the exponents along the time direction are the same, accordingly we have denoted both of them as $\beta_t$ above.
 
The strategy we now follow is similar to the  Type II case. The interpolating metric is given by eq.(\ref{metgen}), which when written out in full becomes
\begin{equation}
\label{intmetVI}
\begin{split}
ds^{2}&=dr^{2}-e^{2\beta_{t} r }dt^{2}\\
&\quad+\left[\left(\frac{1-\tanh\sigma  r }{2}\right) +\left(\frac{1+\tanh \sigma r }{2}\right)e^{2\tilde\beta r }\right]dx^{2} \\
&\quad+\left[\left(\frac{1-\tanh\sigma  r }{2}\right)e^{ 2  \beta_1 r-2 x}+\left(\frac{1+\tanh\sigma  r }{2}\right)e^{2 \tilde\beta r }\right] dy^{2}\\
&\quad+ \left[\left(\frac{1-\tanh\sigma  r }{2}\right)e^{2\beta_2r+2  x}+\left(\frac{1+\tanh \sigma r }{2}\right)e^{2 \tilde\beta r }\right] dz^{2}.
\end{split}
\end{equation} 

As in the Type II case, we again  require that  the interpolating metric  correctly asymptotes  to Type VI in the IR 
and Lifshitz in the UV. This now  imposes the  lower bound
\be
\label{lboundsigVI}
\sigma > \beta_1-\tilde\beta=\beta_2-\tilde\beta.
\ee 
We remind the reader again that the  $r\rightarrow + \infty$ limit is taken  while keeping $x$ fixed to obtain this bound. 
 
We take $\Lambda$ (in $R=1$ units) to have the value   given in eq.(\ref{vallam}). The lower bound for $\sigma $ then becomes,   $\sigma>0.0579912$.
The matter stress tensor is 
then calculated as given in eq.(\ref{fmatterst}) and we examine its properties with respect to the
 energy conditions numerically. 

The numerical analysis shows that 
  as $\sigma$ is increased violations of the null energy condition start setting in around $\sigma= 1.15993$.
The weak energy condition is not violated for smaller values of $\sigma$. 
 Thus, as in the the Type II case, there is a non-vanishing interval for $\sigma$ within which the metric 
has the correct asymptotic behavior and the weak and null energy conditions are both met. 

\begin{figure}
 \begin{center}
 \includegraphics[width=.7\textwidth]{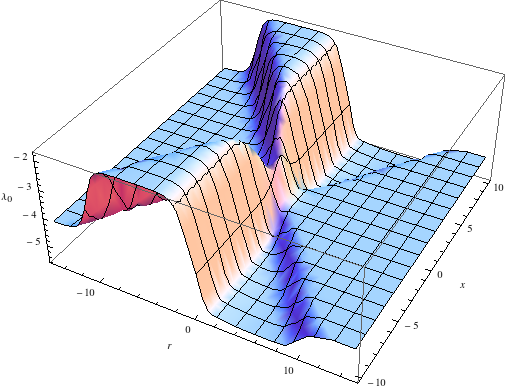}
 \caption{Type VI 3D plot of  $\lambda_0$ (time-like eigenvalue) versus $r$ and $x$    for $\sigma=1,~\Lambda=12$.}\label{6weak}
 \end{center}
 \end{figure}

\begin{figure}
 \begin{center}
  \includegraphics[width=.7\textwidth]{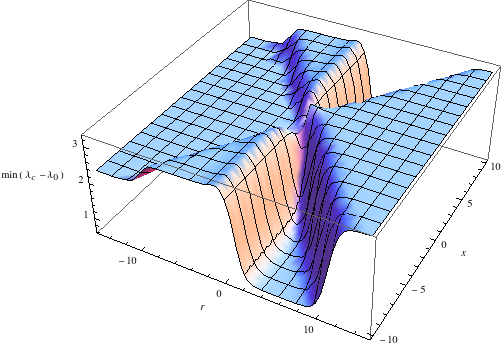}
 \caption{Type VI 3D plot of   $\min(\lambda_{c}-\lambda_0)$ versus $r$ and $x$    for $\sigma=1,~\Lambda=12$.}\label{6null}
 \end{center}
 \end{figure}

\begin{figure}
 \begin{center}
 \includegraphics[width=.7\textwidth]{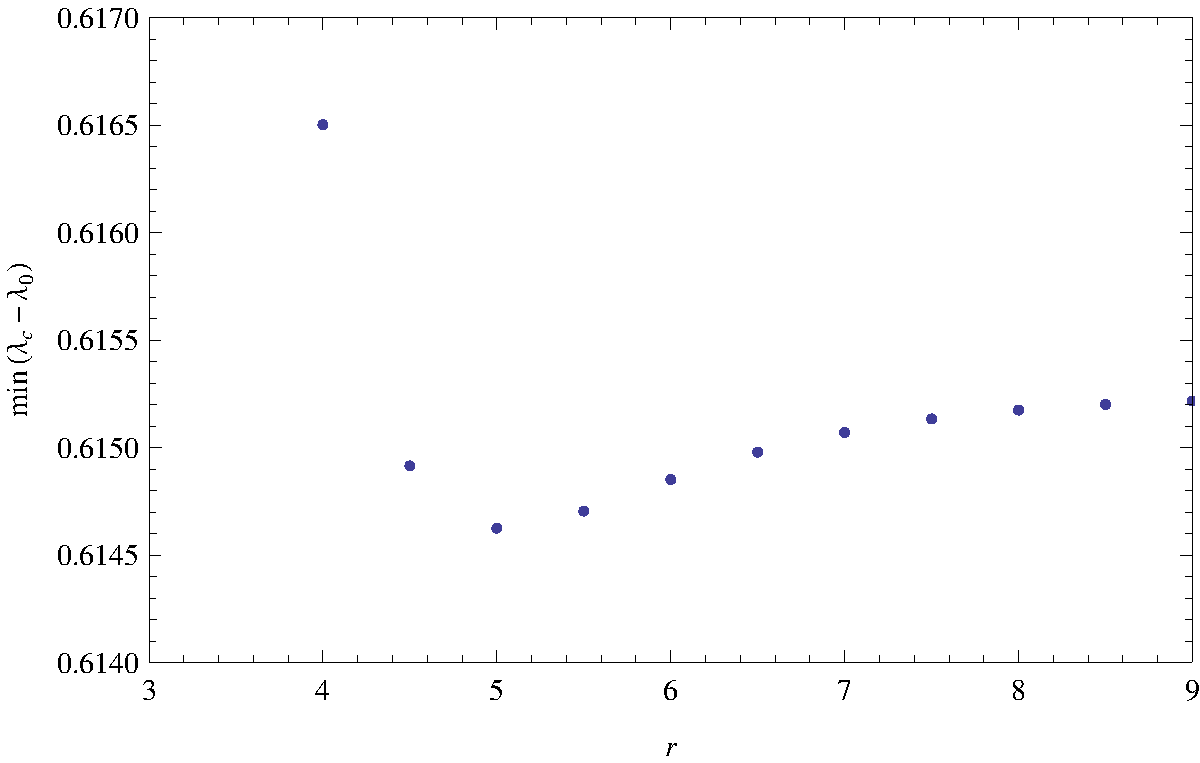}
 \caption{Type VI list plot of   $\min(\lambda_{c}-\lambda_0)$ versus $r$ as $x\rightarrow\pm\infty$ for  $\sigma=1,~\Lambda=12$.}\label{VIxinf}
 \end{center}
 \end{figure}

To illustrate this, consider the case when   $\sigma =1$, which lies within this interval. 
The minimum of the  eigenvalues  of the spatial eigenvectors turns out to be $\lambda_2$, where the eigenvalues are defined 
in eq.(\ref{eig0})--eq.(\ref{eig4}). 
The plots of $\lambda_0$ and $\min(\lambda_c-\lambda_0)=\lambda_2-\lambda_0$, are given in fig. \ref{6weak}, \ref{6null}, 
as a function of the $r, x$ coordinates. We see that the qualitative behavior is similar to that in Type II. $\lambda_0$ is always negative.
And $\lambda_2-\lambda_0$ is positive but there is a worrisome region around $r=5$ where this difference of eigenvalues 
becomes small.  We have analyzed this region  more carefully further. One finds that for any fixed $r \in [4,9]$ the
 minimum value for  $\lambda_2-\lambda_0$ is attained as $|x|\rightarrow \infty$ and moreover this minimum value is positive.
 An analytic expression for this minimum value was also obtained and agrees with the numerical results.  This is shown in fig. \ref{VIxinf} where this minimum value is plotted as a function of $r$ and shown to be positive. 
These results establish that the interpolating metric eq.(\ref{intmetVI}) satisfies both the weak and the null energy conditions when $\sigma$ takes  values within a suitable range.  
 
%

\subsection{Type III}
Since the analysis follows that of the Type VI case closely we will be more brief for this case. 

The invariant one-forms for  Type III, see Appendix A of \cite{IKKNST}, are given by 
\be
\omega^1 = e^{-  x}dy,\quad \omega^2 = dz,\quad \omega^3 = dx.
\ee

Solutions of Type III for the system described by the action  eq.(\ref{actionII}) exist and have been discussed in section 4.2.2 of \cite{IKKNST}.  These  take the form 
eq.(\ref{metVIa}), eq.(\ref{gfII}) for the metric and gauge field. The exponents $\beta_t,\, \beta_1,\,\beta_2$, the  gauge field
$A_t$ and $m^2$ (in $R=1$ units)  are given by 
\begin{align}
\beta_t &= v,\\
\beta_1 &=\beta_3 = 0,\\
\beta_2 &=  \frac{(-2+\Lambda-u)v}{6 - 2\Lambda},\\
m^{2}& =  \frac{1}{2}(4-\Lambda +u),\\
A_{t} & =  \frac{-4v^2  + 2u }{3 -\Lambda },
\end{align}
where
\begin{align}
u&=\sqrt{-8  + \Lambda^2},\\
v&=\frac{\sqrt{-8  + 3 \Lambda + 
  u}}{2} .
\end{align}
 Demanding that $ A_{t},m^{2},\beta_{t},\beta_2 $ be positive and $u$ to be real, we get $\Lambda > 3$.
To obtain the desired interpolation from a Bianchi Type III solution to Lifshitz,
 we follow the strategy  adopted in case of Type II, VI,  above, and consider the following  interpolating metric: 
 \begin{equation}
\label{intmetIII}
\begin{split}
ds^{2}&=dr^{2}- e^{2\beta_{t} r }dt^{2}\\
&\quad+\left[\left(\frac{1-\tanh\sigma  r }{2}\right) +\left(\frac{1+\tanh \sigma r }{2}\right)e^{2\tilde\beta r }\right]dx^{2} \\
&\quad+\left[\left(\frac{1-\tanh\sigma  r }{2}\right)e^{ -2 x}+\left(\frac{1+\tanh\sigma  r }{2}\right)e^{2\tilde\beta r }\right]dy^{2} \\
&\quad+ \left[\left(\frac{1-\tanh\sigma  r }{2}\right)e^{2\beta_2r}+\left(\frac{1+\tanh \sigma r }{2}\right)e^{2\tilde\beta r }\right]dz^{2} .
\end{split}
\end{equation} 

Requiring this interpolating metric to correctly  asymptote to Type VI in the IR and Lifshitz in the UV imposes the
 following lower bound: $\sigma > \beta_2-\tilde\beta$.
We choose   $\Lambda = 12$ in $R=1$ units. 
The lower bound for $\sigma $ then becomes,   $\sigma>0.0456046$.

Furthermore, we numerically  find that violations of the null energy condition start setting in around $\sigma=  0.40108$. 
The weak energy condition is not violated for smaller values of $\sigma$. 
Thus,  we find once again that there is a range of values for $\sigma$ for which the metric asymptotes to the required forms
and for which the weak and null energy conditions are preserved. 

\begin{figure}
 \begin{center}
 \includegraphics[width=.8\textwidth]{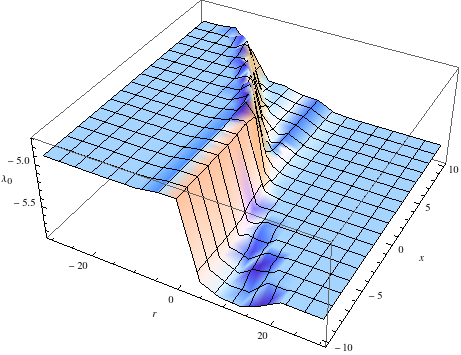}
 \caption{Type III 3D plot of  $\lambda_0$ (time-like eigenvalue) versus $r$ and $x$    for $\sigma=0.3,~\Lambda=12$.}\label{3weak}
 \end{center}
 \end{figure}

\begin{figure}
 \begin{center}
 \includegraphics[width=.8\textwidth]{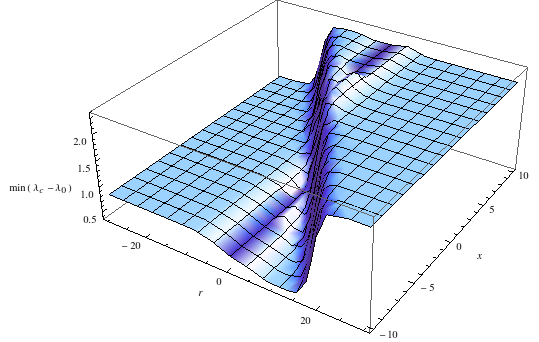}
 \caption{Type III 3D plot of   $\min(\lambda_{c}-\lambda_0)$ versus $r$ and $x$    for $\sigma=0.3,~\Lambda=12$.}\label{3null}
 \end{center}
 \end{figure}

\begin{figure}
 \begin{center}
 \includegraphics[width=.7\textwidth]{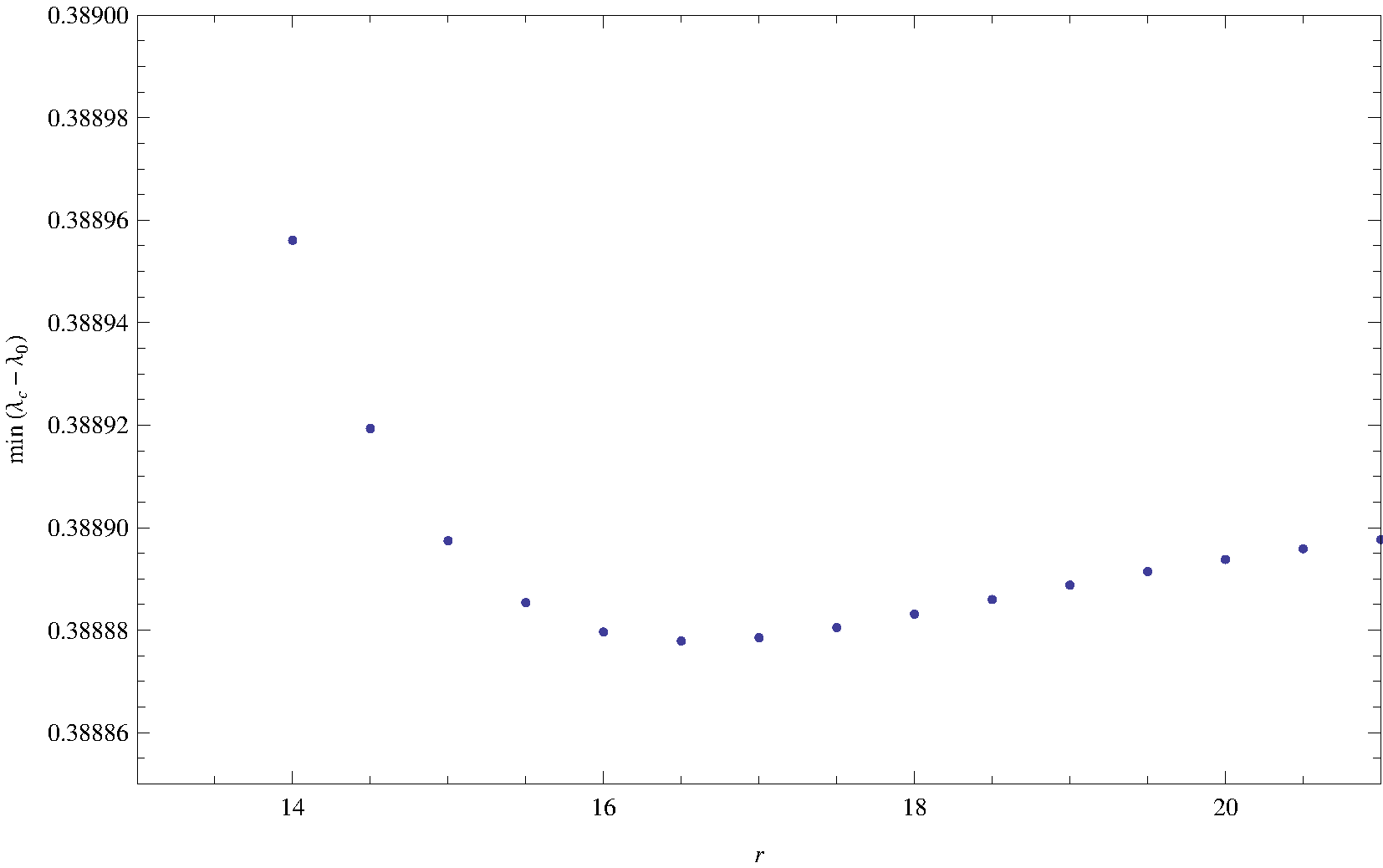}
 \caption{Type III list plot of   $\min(\lambda_{c}-\lambda_0)$ versus $r$ as $x\rightarrow-\infty$ for $\sigma=0.3,~\Lambda=12$.}\label{IIIxinf}
 \end{center}
 \end{figure}

To illustrate this, we choose $\sigma=0.3$ which lies in the allowed region. 
The plots of $\lambda_0$ and $\min(\lambda_c-\lambda_0)=\lambda_2-\lambda_0$, where $\lambda_0$ and $\lambda_2$ are as defined in eq.\eqref{eig0} and eq.\eqref{eig2}, are given in fig. \ref{3weak} and fig. \ref{3null}. 
We see that $\lambda_0$ is always negative. And $\lambda_2-\lambda_0$ is positive but this difference becomes small near 
$r\sim 10-15$ as $x\rightarrow -\infty$. We examined this region in more detail and find that for any fixed $r$ in this region
$\lambda_2-\lambda_0$ attains its minimum value as $x$ is varied for $x\rightarrow -\infty$ and this minimum value is indeed positive.  An analytic expression  for this minimum value was obtained, and agrees with the numerical analysis.
In fig \ref{IIIxinf} we plot  this minimum value, attained when $x\rightarrow -\infty$,  for $\lambda_2-\lambda_0$ against $r$. 
We see that  the minimum value is positive. 
These results establish that the interpolating metric eq.(\ref{intmetIII}) in the Type III case also meets the weak and null energy conditions
for a suitable range of $\sigma$ values.  



\subsection{Type V}
The invariant one-forms in the Type V case are 
\be
 \omega^1 = e^{- x}dy,\quad \omega^2 = e^{- x}dz,\quad \omega^3 = dx.
\ee

Solutions of Type V for the system, eq.(\ref{actionII}) take the form eq.(\ref{metatt}),
eq.(\ref{gfII}). The parameters, $\beta_t,\,\beta_1,\,\beta_2,\, m^2,\, A_t,$ are given by 
\begin{align}
\beta_t &= \sqrt{-4+\Lambda},\\
\beta_1=\beta_2=\beta_3 &=  0, \\
m^{2} &=  0, \\
A_{t} &= \frac{2 (6 -\Lambda)}{4  -\Lambda}.
\end{align}
Demanding that $ A_{t},\beta_{t} $ be positive and real respectively, we get $\Lambda > 6$.
Starting from this metric in the IR one would like to consider a metric of the form eq.(\ref{metgen}) which interpolates to 
Lifshitz space in the UV. However, it turns out that in this case  interpolations of the the  type eq.(\ref{intmetIII}) 
 violate the null energy condition for all values of $\sigma$. 

The failure of the   interpolating metric to work in this case can in fact be understood analytically. 
It is tied to the fact that the Type V solution has one important difference with the other kinds of solutions, 
Type II, VI, III,
studied above. 
Here, it turns out that the smallest eigenvalue of 
$T^{(\mathrm{matter}) \mu}_\nu$ corresponding to a space-like eigenvector, $\min(\lambda_c), c=1,2,3,4$,  is exactly equal
 to the eigenvalue corresponding to the time-like eigenvector, $\lambda_0$,  and thus the  null energy 
condition eq.(\ref{nec2}) is met as an equality. This case is therefore much more  delicate.

In fact,  a perturbative analysis  reveals that once the Type V metric is deformed by considering 
the full interpolating metric given in eq.(\ref{intmetIII}), the splitting which results as $r\rightarrow -\infty$  goes in the wrong direction,
making $\min (\lambda_c) - \lambda_0<0$ for any value of $\sigma$, leading to a violation of the null energy condition.

\section{From Type IX To $AdS_2\times S^3$}
The symmetry algebra for Bianchi Type IX is $SO(3)$ and its natural action is on  a compact space corresponding 
to a squashed $S^3$. 
Therefore, for Type IX it is natural to explore interpolations going from a Type IX attractor geometry to 
$AdS_2 \times S^3$ instead of $AdS_2 \times R^3$ or Lifshitz.

The strategy we use for finding such an interpolation is different from what was used in the cases above. 
It is motivated by the fact that the $SO(3)$ symmetry  for Type IX  is a subgroup of  the symmetries of $S^3$,
$SO(3)\times SO(3)$. The interpolating metric we consider will therefore  be obtained by introducing a deformation parameter which allows  the spatial components  of the metric to go from  that of  a squashed $S^3$ in the IR to the round $S^3$ in the UV. 
This is somewhat akin to  what was done in \cite{IKKNST} to find an interpolation between Type VII and Type I. 

The invariant one-forms for Bianchi Type IX are 
\begin{align*}
 \omega^1&= -\sin(z)\,dx+\sin(x) \cos(z)\,dy,\\
\omega^2&= \cos(z)\,dx+\sin(x) \sin(z)\,dy,\\
\omega^3&=\cos(x)\,dy+dz.
\end{align*}   

One finds that a Type IX  attractor solution  arises  in a system 
of Einstein gravity with
the cosmological constant $\Lambda$, coupled to two gauge fields, $A_1,A_2$ with action
\be
\label{actionIX}
  S= \int d^{5}x \sqrt{-g} \left(R +\Lambda  -\frac{1}{4} F_{1}^{2}-\frac{1}{4} F_{2}^{2}-\frac{1}{4}m^{2}A_{2}^{2}\right).
\ee
Note that   $A_1$ is  massless while
$A_2$ has  $(\mathrm{mass})^2 = m^2$.  

In this solution the  metric  is  given by 
\be
\label{metIXa}
ds^2=R^2[dr^2-e^{2 \beta_t r}dt^2 +(\omega^1)^2+(\omega^2)^2+ \ \lambda \ (\omega^3)^2]
\ee
and the two gauge fields are 
\be
\label{aIX}
A_1=\sqrt{A_t}\, e^{\beta_t r} dt, \hspace{5mm} A_2=\sqrt{A_s}\, \omega^3=\sqrt{A_s}\,(\cos(x)dy+dz).
\ee
 Note that $\lambda$ in eq.(\ref{metIXa}) is the deformation parameter we had mentioned above. 

In $R=1$ units,  the equations of motion which follow from eq.(\ref{actionIX}) give rise to the following relations, 
\begin{eqnarray}
m^2  =  -2\lambda, && A_t=\frac{2(-\lambda +2\Lambda+4)}{-\lambda+2\Lambda +3}, \label{m2IX}\\
A_s=1-\lambda, &&  \beta_t =\left[\frac{-\lambda+2\Lambda+3}{2}\right]^{\frac{1}{2}}\label{asbetaIX}.
\end{eqnarray}
These  relations can be thought of as determining $A_s,A_t, \beta_t, \lambda$ in terms of $\Lambda$ and 
$m^2$. 

Note that the conditions  $A_s, A_t\ge0 ,\,\Lambda>0$ imply, from eq.(\ref{m2IX}) and eq.(\ref{asbetaIX}), the
relation
\be
\label{rel1IX}
\lambda\le1.
\ee
It is easy to see that for $\lambda=1$, this solution becomes\footnote{We note that $(\omega^1)^2+(\omega^2)^2+(\omega^3)^2$ may be obtained as the pullback of the standard Euclidean metric on $R^4$ (with coordinates $W,X,Y,Z$) under the following $S^3$ embedding:
\begin{equation*}
\begin{split}
W = \cos\left(\frac{x}{2}\right)\cos\left(\frac{y+z}{2}\right),&\qquad
X = \cos\left(\frac{x}{2}\right)\sin\left(\frac{y+z}{2}\right),\\
Y = \sin\left(\frac{x}{2}\right)\cos\left(\frac{y-z}{2}\right),&\qquad
Z = \sin\left(\frac{x}{2}\right)\sin\left(\frac{y-z}{2}\right).
\end{split}
\end{equation*}
} $AdS_2\times S^3$, and for any other value 
of $\lambda$ between 0 and 1, it is Type IX.

Let us make one comment before proceeding. Eq.(\ref{m2IX}) and Eq.(\ref{asbetaIX}) give four relations and at first it might seem that they 
determine the four parameters $A_t,A_s, \lambda, \beta_t$ and therefore determine the solution completely. However, since we have set the radius $R=1$, this is not the case and the solution in fact contains one undetermined parameter. 
This becomes clear if we consider the $\lambda=1$ case, where $A_s=0$ and the massive gauge field vanishes. 
The resulting solution is $AdS_2\times S^3$ which is the near-horizon extremal RN solution. This solution has 
one free parameter, which we can take to be  $A_t$, the value of the massless gauge field which determines 
the electric field of this gauge field. Or we can take it to be $R$. 
In the interpolation below, we will take the free parameter to be $R$, and   set  $R=1$,  keeping its value 
 fixed as the radial coordinate $r$ varies.

 
It turns out that for the solution given above 
in eq.(\ref{metIXa}), eq.(\ref{aIX}), eq.(\ref{m2IX}), eq.(\ref{asbetaIX}),  
for any  given  $\lambda$,
 the null energy condition is satisfied but as an equality, with the smallest eigenvalue of a space-like
 eigenvector of $T^{(\mathrm{matter})\mu}_\nu$, $\min (\lambda_c)$,  being equal to the eigenvalue for the 
 time-like eigenvector, $\lambda_0$.  This is analogous to what we saw above  in the Type V case. 
However, here because the symmetries  involved are different, we can choose another  kind of interpolation, as mentioned
in the beginning of this section.

We do this   by taking    $\lambda$ to be   a function of  $r$  of the form 
\be
\lambda (r)= C+ (1-C) \left( {1+\tanh (\sigma r) \over 2}\right)
\ee
where $C$, and $\sigma$ are constants,   with $0<C<1$, to meet eq.(\ref{rel1IX}). We find that the degeneracy
between $\min(\lambda_c), \lambda_0$  is now  lifted. Unlike the Type V case though, 
 this lifting occurs so that $\min(\lambda_c)-\lambda_0>0$, if $\sigma$ is sufficiently small, thus preserving the null energy condition, eq.(\ref{nec2}).
If $\sigma$ becomes bigger than a critical value, violations of the NEC set in. 

For example, for the choice of $\Lambda=12$,  and  $ C = 0.5 $ we find that the energy conditions are met for a range of $\sigma$ 
up to $\sigma_{\mathrm{crit}}=1.82$. 
For  $0 < \sigma \le 1.82 $ and  $ C = 0.5 $ both eq.(\ref{wec2a}), eq.(\ref{wec2b}) are met,
so that the interpolating metric above satisfies the WEC and hence also the NEC. 

\begin{figure}
 \begin{center}
 \includegraphics[width=.7\textwidth]{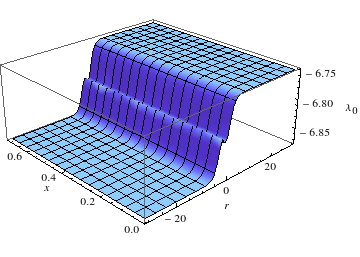}
 \caption{Type IX 3D plot of  $\lambda_0$ (time-like eigenvalue) versus $r$ and $x$    for $C=0.5,~\sigma=0.5,~\Lambda=12$.}\label{9weak}
 \end{center}
 \end{figure}

\begin{figure}
 \begin{center}
 \includegraphics[width=.7\textwidth]{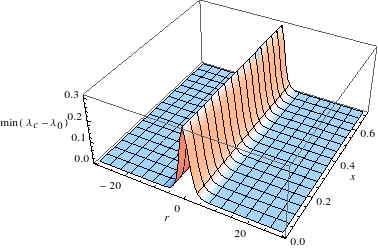}
 \caption{Type IX 3D plot of   $\min(\lambda_{c}-\lambda_0)$ versus $r$ and $x$    for $C=0.5,~\sigma=0.5,~\Lambda=12$.}\label{9null}
 \end{center}
 \end{figure}

For $C=0.5,~\sigma=0.5$,  
the results are  summarized in fig. \ref{9weak}  and \ref{9null}.
Fig. \ref{9weak} shows that $\lambda_0$ 
 satisfies the condition $\lambda_0<0$. And  fig. \ref{9null} shows that 
$\min(\lambda_c-\lambda_0)>0$.
As $r\rightarrow \pm \infty$ the interpolation approaches a solution of the type considered in eq.(\ref{metIXa}), 
eq.(\ref{aIX}), and the 
value of $\min(\lambda_c-\lambda_0) \rightarrow 0$. However, we have verified that at both ends,  $r\rightarrow \pm \infty$,
$\min(\lambda_c-\lambda_0)$ approaches zero from above so that the NEC continues to hold.
Together, these results imply that $T^{(\mathrm{matter}) \mu}_\nu$ satisfies the weak energy condition, and 
therefore also the null energy condition.  

\section{C-function} 

In this section, we investigate a large class of 
  geometries of the form
\be
\label{fint1}
ds^2=-g_{tt}(r) dt^2 + dr^2 + g_{ij}(x^i,r) dx^idx^j
\ee
which interpolate between two Bianchi attractor spacetimes. The Bianchi attractors arise
at the UV and IR ends, $r\rightarrow \pm \infty$ respectively,
  where the geometry takes the scale invariant form, eq.(\ref{metatt}), with the exponents $\beta_t, \beta_i$ being constant and positive. 
The UV and IR ends  are defined by the redshift factor, $g_{tt}$, which 
 decreases  from the  UV to the  IR. 
  
We find that as long as the matter sourcing the geometry satisfies the null energy condition, the area 
element  of the  submanifold spanned by the $x^i$ coordinates (at constant $t,r$) monotonically decreases with $r$, 
obtaining its minimum value in the IR.
For a Bianchi attractor, eq.(\ref{metatt}), the area element is proportional to $e^{\sum_i \beta_i r}$ and diverges in the UV,  $r\rightarrow \infty$,
 while vanishing in the IR, $r\rightarrow -\infty$. The only exception is when the exponents $\beta_i$ all 
vanish, as happens for example in  $AdS_2\times R^3$ space, in which case the area 
element becomes a non-zero constant. 
 We also find an additional function, which we will refer to as the C-function below,  which is monotonically decreasing from the UV to the IR. For an AdS attractor,  this function attains a constant value and is the 
central charge. For other Bianchi attractors meeting a specific condition, given in eq.(\ref{condcba})
 below, 
this function also flows to a constant in the near-horizon region.  More generally,  when this  specific condition is not met,
 the function either vanishes or diverges as $r\rightarrow \pm \infty$. All of these  results are most easily derived  by applying Raychaudhuri's equation to  an appropriately chosen set of null  geodesics in the geometry, eq.(\ref{fint1}). 

Note that the flows we study include interpolations between two AdS spacetimes which at 
intermediate values of $r$ can break not only Lorentz invariance but also spatial rotational 
invariance and translational invariance. As long as the UV and IR geometries are AdS, our results imply 
that the IR central charge must be smaller than the UV one. Our results  therefore lead to  
a generalization of the holographic C-theorem for flows between conformally invariant theories which can
also break boost,  rotational and translational symmetries. 
This is in contrast to much  of the discussion in the literature so far, 
 which has considered   only Lorentz invariant flows. 

Besides the area element and the C-function  mentioned above, and of course  monotonic functions of these, 
for example, powers of the area element or the  C-function, we do not find any other function
 which in general would 
 necessarily be monotonic as a consequence of the null energy condition. As was mentioned above, 
both the area element  and the  C-function  do not in  general  attain 
 finite non-vanishing values in the asymptotic Bianchi attractor regions. This suggests that for 
 Bianchi attractors in general, no analogue of a finite, non-vanishing, central charge  can be defined which is monotonic under RG flow. 
This conclusion should apply for example to general  Lifshitz spacetimes (see also a discussion of these cases in \cite{Liu}). 
When the Bianchi attractor meets the  specific condition of eq.(\ref{condcba}), the C-function 
 does become a 
finite constant and the analogue of the central charge can   be defined. 
Understanding this constant  in the field theory dual to 
 the Bianchi attractor  would be a  worthwhile thing  to do. 

\subsection{The Analysis}

We now turn to describing the analysis in more detail. Our notation will follow that of \cite{Wald}, Section 9.2.
The analysis is also connected to the discussion of a C-function in \cite{GJMT}. A nice discussion of the C-function in AdS space can be found in Section 4.3.2 of \cite{MAGOO}. For discussions of renormalization group flows
in the context of the AdS/CFT correspondence, see 
\cite{Akhmedov:1998vf,Alvarez:1998wr,Gorsky:1998wn,Porrati:1999ew,Balasubramanian:1999jd}.  The earliest proofs of
holographic C-theorems appear in \cite{Freedman:1999gp}, \cite{Girardello:1998pd}, and our strategy is a generalization of 
the one employed there.


We start with a spacetime described by the metric, eq.(\ref{fint1}), 
and consider a $3$-dimensional submanifold 
spanned by the $x^i$  coordinates  for any fixed $r,t$.    
Next, we   consider a  family of null geodesics which emanate from all points of this submanifold.
If $n^a$ is the tangent vector of the null geodesic, with $a$ taking the values
$a=t,r,i=1,2,3$, then the geodesics we consider have $n^i=0$ so that they  correspond to motion only 
in the radial direction. Both the radially in-going and out-going families of this type
form a congruence. To arrive at our results, it is enough to consider any one of them and we consider
 the radial out-going geodesics below. 
The time-like component of the vectors in this congruence, 
$n_t$, is a constant which we can set to unity, 
\be
\label{valnt}
n_t=1.
\ee
Then for the radially outgoing geodesics 
\be
\label{valnr}
n^r= {dr \over d\lambda}={1\over \sqrt{g_{tt}}},
\ee
where $\lambda$ is the affine parameter along the geodesic.

 Now we take   the tensor field
\be
\label{defb}
B_{ab}=\nabla_b n_a
\ee
and consider its components for $a,b=i,j=1,2,3$. In the notation of \cite{Wald}, 
this gives us the components of $\hat{B}_{ab}$. 
It is easy to see that
\be
\label{antb}
B_{ij}=-\Gamma^c_{ij} n_c = {1\over 2} {\partial_rg_{ij} \over g_{tt}}
\ee
and thus $B_{ij}$ is symmetric so that the twist of the congruence vanishes.  
The expansion of the congruence, denoted by $\theta$, is then 
\be
\label{valtheta}
\theta ={1\over 2} \partial_r g_{ij} {g^{ij}\over \sqrt{g_{tt}}} = \partial_r(\ln A){1\over 
\sqrt{g_{tt}}},
\ee
where we have introduced the notation
\be
\label{defA} A\equiv \sqrt{\det(g_{ij})}
\ee
to denote the area element of the hypersurface spanned by the $x^i$ coordinates for any constant $r,t$.
 
From eq.(\ref{valtheta}) and eq.(\ref{valnr}) we get that  
\be
\label{dervt}
{d \theta \over d\lambda}={1\over \sqrt{g_{tt}}} \partial_r \left({\partial_r\ln A \over \sqrt{g_{tt}}}\right).
\ee
Raychaudhuri's equation then gives
\be
\label{re}
{d \theta \over d\lambda} = -{1\over 3} \theta^2-\hat{\sigma}_{ab}\hat{\sigma}^{ab}-R_{cd}n^cn^d
\ee
since the twist $\hat{\omega}_{ab}=0$. Note that the coefficient of the first term on the RHS is 
${1\over 3}$ and not ${1\over 2}$ since we are in $4+1$ dimensions and not $3+1$ dimensions. 
 
If the matter sourcing the geometry satisfies the null energy condition, the Ricci curvature satisfies the relation $R_{cd} n^cn^d\ge 0$,  leading to the conclusion from eq.(\ref{re}) that 
${d\theta \over d\lambda}<0$. From eq.(\ref{dervt}), this in turn  leads to 
\be
\label{derva}
\partial_r\left({\partial_r\ln A \over \sqrt{g_{tt}}}\right) <0.
\ee
In the UV, $r\rightarrow \infty$,
\be
\label{uvb}
{\partial_r\ln A \over \sqrt{g_{tt}}}=\sum_i\beta_i e^{-\beta_t r} >0
\ee
where $\beta_i,\beta_t$ are the exponents corresponding to the UV attractor. 
It then follows from eq.(\ref{derva}) that for all values of  $r$, 
 ${\partial_r\ln A \over \sqrt{g_{tt}}}>0$,
and thus 
\be
\label{dervac}
\partial_r \ln A>0.
\ee
This leads to our first result:  the area element $A$, defined in eq.(\ref{defA}), decreases monotonically
from the UV, $r\rightarrow \infty$, to the IR, $r\rightarrow -\infty$.

Raychaudhuri's equation, eq.(\ref{re}) also leads to the conclusion that 
\be
\label{rec2}
{d\theta\over d\lambda}+{1\over 3} \theta^2 \le 0,
\ee
if the matter satisfies the null energy condition. 
From eq.(\ref{valtheta}), eq.(\ref{dervt}) this leads to 
\be
\label{secres1}
\partial_r \left( { (\partial_r \ln A)  A^{1/3} \over \sqrt{g_{tt}}} \right)<0.
\ee
A monotonically decreasing function from the UV to the IR is therefore   given by 
\be
\label{cdef}
C=\left({\sqrt{g_{tt}} \over (\partial_r\ln A) A^{1/3} }\right )^3.
\ee
For a Bianchi attractor with exponents $\beta_t,\beta_i$, $C$ becomes
\be
\label{valcba22}
C \propto \left({e^{(\beta_t-\bar{\beta}) r} \over 3  \bar{\beta}}\right)^3,
\ee
where we have defined 
\be
\label{defbarb}
\bar{\beta}={1\over 3} \sum_i \beta_i.
\ee
The overall power of $3$ in the definition of $C$, eq.(\ref{cdef}),  is chosen so that in AdS space, 
 where   $\beta_i=\beta_t$ and   $C$ is a constant, it agrees  with the usual 
definition of the central charge up to an overall coefficient. More generally,  $C$ also becomes a constant 
 for any Bianchi attractor meeting the condition 
\be
\label{condcba}
\beta_t=\bar{\beta} = {1\over 3} \sum_i \beta_i
\ee
and   now takes a   value 
\be
\label{cvalspba}
C \propto {1\over \left(\sum_i \beta_i\right)^{3}}.
\ee
However, for the general case of a Bianchi attractor which does  not meet the condition in 
eq.(\ref{condcba}),   $C$ does not attain a constant value. In such situations, for  $C$ to be monotonically decreasing towards the IR or constant, we need $(\beta_t-\bar{\beta}) \ge 0 $. Thus, we find that if the attractor arises in the IR, then  our $C$  vanishes. On the other hand, if the attractor arises in the UV, it diverges.

\section{Conclusions}

In this paper, we constructed a class of smooth metrics which interpolate from various Bianchi attractor
 geometries in the IR to Lifshitz spaces or $AdS_2\times S^3$ in the UV. We did not show that  these    
 interpolating metrics   arise as solutions to Einstein gravity coupled with suitable matter field theories. However, 
for Bianchi Types II, VI (with parameter $h=-1$), III and IX, we did show  that {\it were}  these geometries 
 to arise as solutions to Einstein's equations, the required  matter would  not violate the weak or 
null energy conditions. It is well known that the  Lifshitz spaces (which are in fact  attractors of Bianchi Type I) or $AdS_2\times S^3$ geometry in turn can be connected to  $AdS_5$ in the ultraviolet, 
 with no non-normalizable deformation for the metric being turned on in the asymptotic $AdS_5$ region. 
Thus, our results establish that there is no barrier, at least at the level of energy conditions,
 to having a smooth interpolating metric arise as a solution of the Einstein equations sourced by  reasonable matter,
 which connects the various Bianchi classes mentioned above with asymptotic $AdS_5$ space.
We should mention here that for Type VII geometries, which were not investigated in this paper, 
 solutions with reasonable matter which interpolate from the attractor region
 to $AdS_2\times R^3$ or $AdS_5$ are already known to exist \cite{IKKNST}.

The absence of any non-normalizable metric deformations in the asymptotic $AdS_5$ region in our interpolations
 suggests that the Bianchi attractor geometries can arise as the dual description in the IR of field theories 
which live  in flat space. The anisotropic and homogeneous phases in these field theories, 
described by the Bianchi attractor regions, could arise either due to a spontaneous breaking of 
rotational invariance or due to its breaking by sources other than the metric in the field theory.  
We expect both possibilities to be borne out. For spin density waves, which correspond to Type VII, indeed this is already known to be true \cite{gauntlett,IKKNST}.

Finding such interpolating metrics  as solutions   to Einstein's equations is not easy, as was mentioned in the introduction, since it requires solving coupled partial differential  equations in at least two variables. 
We hope that the  results presented  here will provide some further motivation to try and address this challenging problem.  
Perhaps it might be best to first look for supersymmetric domain walls interpolating between
different Bianchi types, since  for such solutions, working with first-order equations often   suffices.

We also note that our smooth interpolating metric which interpolates from Bianchi Type V to Lifshitz failed to satisfy the null energy conditions. 
Our failure in this case may be due to the restricted class of functions we used to construct the  interpolating metrics or 
it might suggest a more fundamental constraint. We leave a more detailed exploration of this issue for the future.

Towards the end of the paper, we explored whether a C-function exists for flows between two Bianchi attractor geometries. As long as the matter sourcing the geometry
meets the null energy condition, we found that a function can be defined which is monotonically decreasing from the UV to the IR. In
 AdS space, this function becomes the usual central charge. More generally though, unless the Bianchi 
attractor meets a specific condition relating the exponents $\beta_i,\beta_t$ which characterize it, 
the function we have identified does not attain a finite, non-vanishing constant value in the attractor
 geometry.  The absence of a general monotonic 
function  which is non-vanishing and  finite  in  the attractor spacetime suggests that no
 analogue of a central charge, which is monotonic under RG flow, 
 can be defined in general for field theories dual to the Bianchi attractors. 
For flows between AdS spacetimes, on the other hand,  our analysis implies that the central charge decreases even under RG flows which break boost, rotational and translational invariance.  

\bigskip
\centerline{\bf{Acknowledgements}}
\medskip

We thank Sarah Harrison, Nori Iizuka, Prithvi Narayan, Ashoke Sen,  Nilanjan Sircar and Huajia Wang for discussions. S.P.T. acknowledges support from the J.~C. ~Bose Fellowship, Department of Science and Technology, Government of India. S.K. and  S.P.T. acknowledge the hospitality of the
Simons Symposium on Quantum Entanglement, where some of this research was carried out. N.K., R.S. and S.P.T.
 acknowledge support from the Government of India and most of all are grateful to the people of India
for generously supporting research in string theory. A.S. would like to thank the Department of Physics, Indian Institute of Technology -- Bombay, for its support. S.K. is supported by the National Science Foundation under grant
no. PHY-0756174, the Department of Energy under contract DE-AC02-76SF00515, and the John Templeton Foundation.
   
\appendix

\section{The Weak and Null Energy Conditions}

We shall now review the weak and null energy conditions in detail. The weak energy
condition (WEC) stipulates that the local energy density as observed by a time-like observer is nonnegative. In other 
words, if $u^\mu$ are the components of a time-like vector, we must have $T_{\mu\nu}u^\mu u^\nu \ge 0$ everywhere, with $T_{\mu\nu}$ being the components of the stress tensor. Note that if we raise one of the indices of $T_{\mu\nu}$ to get $T^\mu_\nu$, we could interpret the stress tensor as a linear transformation $T$ that acts on the components of a vector $u$ via $(Tu)^\mu=T^\mu_\nu u^\nu$.

The WEC now simply becomes $\langle u,Tu\rangle\ge 0$, where the angle brackets denote the inner product with respect to the metric. Since $T$ is a linear transformation from a vector space to itself, it makes sense to talk of the eigenvalues and eigenvectors of $T$. In particular, if $u$ is a time-like eigenvector which is normalized so that $\langle u,u\rangle =-1$ and which  belongs to some eigenvalue $\lambda$ (not to be confused with the $\lambda$ parameter we had introduced in the interpolating metric), then we have
\begin{equation}
\langle u,Tu\rangle=\lambda\langle u,u\rangle=-\lambda.
\end{equation}
Thus, a necessary condition for the WEC to hold is that the eigenvalues corresponding to all time-like eigenvectors of $T$ be non-positive.

Note that this isn't a sufficient condition for the WEC to hold.
To go further, let us first note that 
   $T$ is self-adjoint:
\begin{equation*}
\langle u, Tv\rangle=T_{\mu\nu}u^\mu v^\nu = \langle Tu,v\rangle.
\end{equation*} 
However, it does not follow from this property that $T$ is diagonalizable and that its eigenvalues are necessarily real, since the inner product is indefinite in a Lorentzian metric. 
For the metrics we deal with in the paper, we fortunately do not have to deal 
 with this complication because, it turns out that in all the cases we analyze,
 $T$ does turn out to be diagonalizable with real eigenvalues. Accordingly, we restrict our discussion to this case below.

It then follows that there exists a vierbein $\{u_0,u_1,u_2,u_3,u_4\}$ consisting of the eigenvectors of $T$, which is orthonormal in the sense that $\langle u_a,u_b\rangle=\eta_{ab}$. 
If we let $Tu_a = \lambda_au_a$, then our claim is that the WEC is equivalent to the following statement: $\lambda_0\le0$ and $\lvert \lambda_0\rvert +\lambda_c \ge 0$ for $c=1,2,3,4$.

To prove necessity, we note that we have already shown that $\lambda_0\le 0$. Now, for an arbitrary
time-like vector of the form $v=Au_0+Bu_c$, where $c$ can be 1, 2, 3 or 4, we have $\langle v,v\rangle=-A^2+B^2<0$. By the WEC we have
\begin{equation*}
\langle v,Tv\rangle=|\lambda_0|A^2+\lambda_cB^2\ge0.
\end{equation*} 
If we let $\epsilon = A^2 -B^2$, the above can rewritten as
\begin{equation*}
(|\lambda_0|+\lambda_c)B^2+\epsilon|\lambda_0|\ge0.
\end{equation*}
Since $v$ is arbitrary, $\epsilon$ can be an arbitrarily small positive real number. It follows that $\lvert \lambda_0\rvert +\lambda_c \ge 0$ for $c=1,2,3,4$.

To prove sufficiency, we note that a generic time-like vector $v$ may be given by 
\begin{equation*}
v = Au_0+Bu_1+Cu_2+Du_3+Eu_4.
\end{equation*}
where the coefficients are subject to the following
\begin{equation*}
A^2>B^2+C^2+D^2+E^2.
\end{equation*}
The conditions $\lambda_0\le0$ and $\lvert \lambda_0\rvert +\lambda_c \ge 0$ for $c=1,2,3,4$ hence guarantee that
\begin{align*}
\langle v,Tv\rangle&=|\lambda_0|A^2+\lambda_1B^2+\lambda_2C^2+\lambda_3D^2+\lambda_4E^2\\
&\ge|\lambda_0|(B^2+C^2+D^2+E^2)+\lambda_1B^2+\lambda_2C^2+\lambda_3D^2+\lambda_4E^2\\
&=(\lvert \lambda_0\rvert +\lambda_1)B^2+(\lvert \lambda_0\rvert +\lambda_2)C^2+(\lvert \lambda_0\rvert +\lambda_3)D^2+(\lvert \lambda_0\rvert +\lambda_4)E^2\\
&\ge 0.
\end{align*}
In fact, we can go further and easily show this implies the null energy condition (which states that $\langle n,Tn\rangle\ge 0$ for all 
null vectors $n$ everywhere) by following the same outline as the proof above. We note that a generic null vector $n$ may be given by
 \begin{equation*}
n = Au_0+Bu_1+Cu_2+Du_3+Eu_4,
\end{equation*}
where the coefficients are subject to the following
\begin{equation*}
A^2=B^2+C^2+D^2+E^2.
\end{equation*}
The conditions $\lambda_0\le0$ and $\lvert \lambda_0\rvert +\lambda_c \ge 0$ for $c=1,2,3,4$ hence guarantee that
\begin{align*}
\langle n,Tn\rangle&=|\lambda_0|A^2+\lambda_1B^2+\lambda_2C^2+\lambda_3D^2++\lambda_4E^2\\
&=\lambda_0|(B^2+C^2+D^2+E^2)+\lambda_1B^2+\lambda_2C^2+\lambda_3D^2+\lambda_4E^2\\
&=(\lvert \lambda_0\rvert +\lambda_1)B^2+(\lvert \lambda_0\rvert +\lambda_2)C^2+(\lvert \lambda_0\rvert +\lambda_3)D^2+(\lvert \lambda_0\rvert +\lambda_4)E^2\\
&\ge 0,
\end{align*}
which is the null energy condition (NEC). 
Thus, in terms of the eigenvalues, the NEC is equivalent to the following statement:  $ - \lambda_0 +\lambda_c \ge 0$ for $c=1,2,3,4$ where $\lambda_0$ is 
the eigenvalue corresponding to the time-like eigenvector and $\lambda_c$ corresponds to any of the space-like eigenvectors. 

To summarize the above observations:
\begin{enumerate}
 \item For the WEC, it suffices to have (i) $\lambda_0\le 0$  and (ii) $\lvert \lambda_0\rvert +\lambda_c \ge 0$ for $c=1,2,3,4$.
\item  For the NEC, it suffices to have  $ - \lambda_0 +\lambda_c \ge 0$ for $c=1,2,3,4$, where $\lambda_0$ is the eigenvalue corresponding to the 
time-like eigenvector and $\lambda_c$ corresponds to any of the space-like eigenvectors.
\end{enumerate}


\end{document}